\documentclass[conference]{IEEEtran}

\usepackage{ctable}
\usepackage{cite}
\usepackage[cmex10]{amsmath}
\usepackage{graphicx}
\usepackage{multirow}
\usepackage{listings}
\usepackage{color}
\usepackage{xcolor}
\usepackage{balance}
\usepackage{algorithm2e}

\begin{document}

\title{Enhanced Red-Black-Tree Data Structure for Facilitating the Scheduling of Reservations}

\author{\IEEEauthorblockN{Marcos Dias de Assun\c{c}\~ao}
\IEEEauthorblockA{Inria, LIP Laboratory\\
Ecole Normale Sup\'{e}rieure de Lyon\\
University of Lyon, France\\
assuncao@acm.org}
}

\maketitle

\begin{abstract}
This paper details a data structure for managing and scheduling requests for computing resources of clusters and virtualised infrastructure such as private clouds. The data structure uses a red-black tree whose nodes represent the start times and/or completion times of requests. The tree is enhanced by a double-linked list that facilitates the iteration of nodes once the start time of a request is determined by using the tree. We describe the data structure main features, provide an example of use, and discuss experiments that demonstrate that the average complexity of two operations are often below 10\% of their respective theoretical worst cases.
\end{abstract}

\IEEEpeerreviewmaketitle

\section{Introduction}
\label{sec:introduction}

Advances in IT have led to the emergence of virtualisation and provisioning models where resources are provided to client applications on demand. Under such models, often termed as cloud computing \cite{ArmbrustCloud:2009}, customers request resources to run their applications and pay only for what they consume. Techniques to manage the allocation of resources and schedule user requests are often at the core of cloud management systems. 

Although on-demand provisioning was initially the only model offered by clouds, other approaches such as reserved virtual machines\footnote{http://aws.amazon.com/ec2/purchasing-options/reserved-instances/} and spot instances\footnote{http://aws.amazon.com/ec2/purchasing-options/spot-instances/} have later gained popularity. Resource reservations are of interest to users as they provide means for reliable allocation and enable users to plan the execution of their applications. Previous work in large-scale computing infrastructure demonstrates that certain deadline-constrained applications demand predictable quality of service \cite{SinghHPDC:2007}, often requiring a number of computing resources to be available over a well defined period, commencing at a specific time in the future; good requirements for advance reservation. 

For scheduling decisions, management systems generally maintain information on resource availability in data structures or databases~\cite{JacksonMAUI:2001}. The systems may need to handle numerous requests per minute, with each request arrival or completion triggering scheduling operations requiring multiple reads or updates to the data structure. Efficient data structures are essential to timely check whether reservations or ordinary requests can be accommodated, or to provide alternatives to users with flexible requests \cite{RoblitzARPlacement:2006}. Several of these operations are generally referred to as \textit{admission control}.

This paper describes a data structure for storing information on computing resource availability and performing admission control of ordinary requests and reservations of computing resources. The data structure uses a red-black tree; a binary search tree with one additional attribute per node: its colour, which can be either red or black~\cite{CormenIntroAlgo:2001}. The structure is enhanced by a double linked list used to iterate nodes that contain the resource availability when checking whether a request can be accepted. The data structure, termed as ``Availability Profile", or ``Profile" for short, maintains information on resources available when requests start or complete.

The rest of this paper is organised as follows. Section~\ref{sec:related_work} describes background and related work. In Section~\ref{sec:availability}, we introduce the data structure, whereas Section~\ref{sec:using_profile} illustrates how it can be used to build scheduling policies. Section~\ref{sec:experiments} contains results on evaluating the practical average complexity of two operations, and Section~\ref{sec:conclusions} concludes the paper.


\section{Background and Related Work}
\label{sec:related_work}

\subsection{Resource Reservations}

Although initially oriented towards on-demand provisioning, cloud computing solutions have later introduced other means to provide resources to client applications, including frameworks that enable advance and immediate resource reservations\footnote{https://wiki.openstack.org/wiki/Blazar}. In the past, other systems have benefitted from reservations, including grids where large-scale experiments can demand co-allocation of resources across several sites \cite{BoghosianGridsofGrids:2006}.

Systems that manage request scheduling and resource allocation, generally employ a data structure to store information on resources available until a particular time in the future. The structure is examined to check whether a request can be admitted or not. This period over which the availability information is stored depends on the resource allocation policy in use. For example, for an allocation policy that schedules requests using conservative backfilling~\cite{MuAlemBackfilling:2001} and reserves resources as requests arrive, this period may vary according to the number of requests currently in the system. Under aggressive backfilling~\cite{LifkaBackfilling:1995}, this period is often shorter as the scheduler maintains details on running requests and about the first waiting request.

\subsection{Data Structures Using Slotted Time and Non-Slotted Time}

In slotted-time data structures a period over which the availability information is divided in time frames of equal length~\cite{BurchardDataStructure:2005,SulistioGarQ:2008}. If accepted, a request is allocated a number of consecutive slots for a period long enough to accommodate it (\textit{i.e} a number of slots whose total duration is equal or greater than time frame initially requested).

The data structure presented here does not use slotted time, thus allowing for finer time granularity for accepted requests as the duration of allocated time frames does not need to be multiple of slot length. As discussed next, the structure uses ranges of available resources as it needs to ensure that the same resources are allocated to a request during its whole execution. With slotted time and short slots, the profile would have this range information replicated at all time frames, and iterating slots would be time consuming.


\section{The Availability Profile}
\label{sec:availability}

The proposed data structure follows the concept of availability profile~\cite{MuAlemBackfilling:2001}, which utilises a list whose entries contain information about the ranges of resources available after start and completion of requests. This paper enhances the concept of availability profile by:

\begin{itemize}
\item allowing it to maintain information about reservations;
\item using a red-black tree to search for the start of a free time frame suitable for scheduling a request, thus reducing the complexity from $O(n)$ when using a sorted list to $O(log\ n)$ by using the tree; and
\item storing information about the ranges of resources available at each node, hence enabling various policies to select time frames, such as first-fit, best-fit and worst-fit.
\end{itemize}

\begin{figure}[ht]
\centering 
\includegraphics[width=.95\columnwidth]{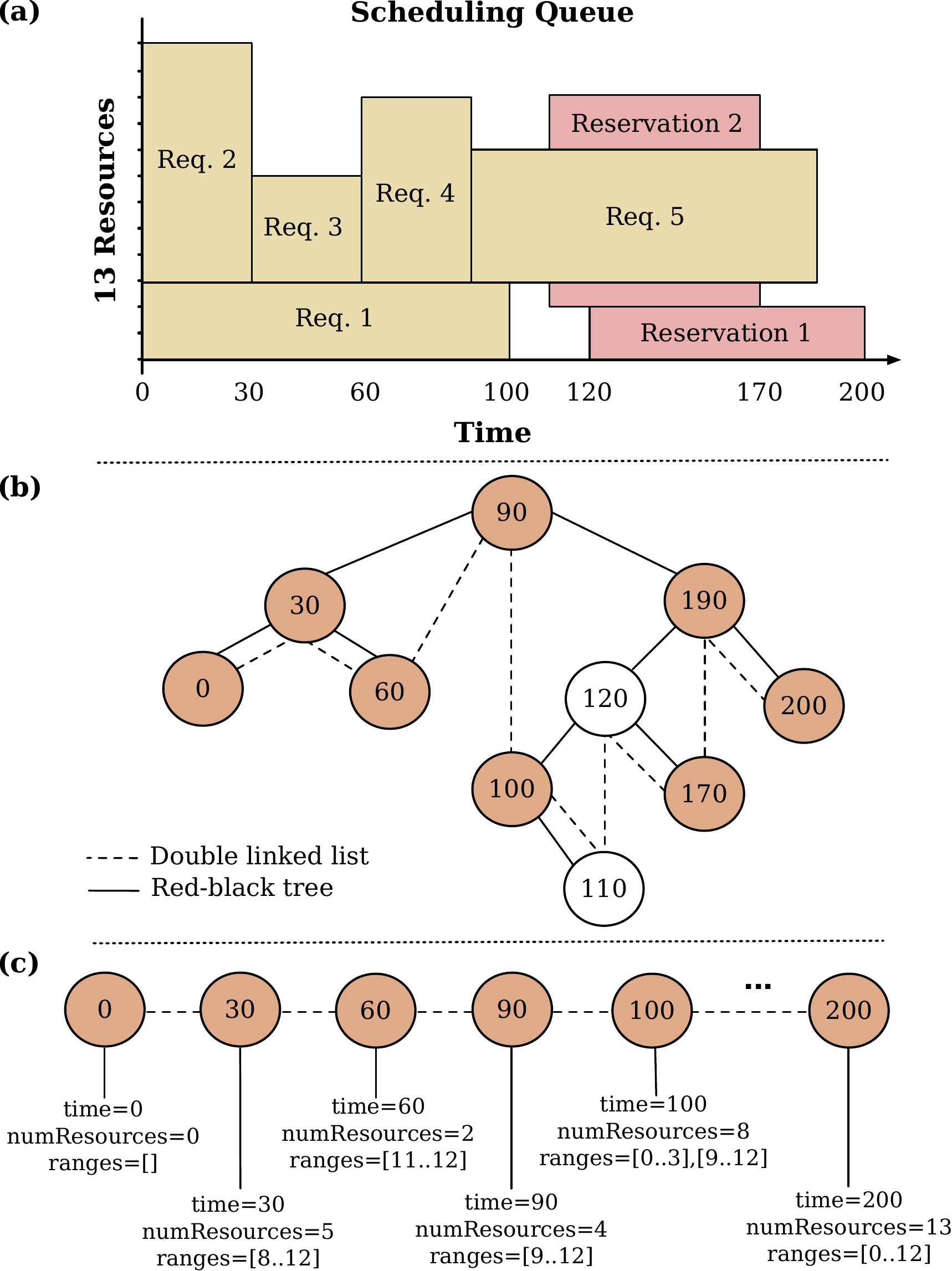} 
\caption{Pictorial view of the data structure: (a) scheduling queue of a cluster with 13 resources; (b) availability information as a red-black tree; and (c) information stored by the nodes.}
\label{fig:tree_example}
\end{figure}

A red-black tree is approximately balanced due to the manner nodes are coloured from the root to a leaf, which ensures that no path is more than twice as long as any other. After modifying a red-black tree, rotation and colour change operations guarantee that it remains approximately balanced. The nodes of the tree contain information about resources available at specific times in future; the start and/or completion times of requests. The profile utilises ranges of resources as it needs to know whether the selected resources would be available over the entire period requested. For instance, a cluster with 10 resources has a range from 0 to 9 (\textit{i.e.} [0..9]). This contrasts with data structures that store bandwidth available on a network link, as in the latter the availability at a time is generally a single number~\cite{BurchardDataStructure:2005}. The profile needs to ensure that the same resources are available over the period requested as starting a request on a set of resources and migrating it several times during execution is undesirable. Here a resource represents a slot (\textit{i.e.} a combination of number of vCPUs, memory and storage) to run a virtual machine, but the structure is generic enough to allow a scheduler to work with other types of abstractions. 

The red-black tree is used to locate the node that represents the start of a request, termed as anchor, whereas the double-linked list is employed to iterate nodes once the anchor is found. By using the list, all nodes until the supposed request completion time are verified to check whether there are resources available to admit the request into the system. We provide an example of a cluster of 13 resources to depict how the data structure works (see Figure~\ref{fig:tree_example}). Figure~\ref{fig:tree_example}~(a) shows the scheduling queue at time 0 --- the queue contains both \textit{best-effort requests} and \textit{reservations}. Although a best-effort request starts as soon as enough resources are available, a reservation requires resources over a well-defined time frame. The operations for obtaining a time frame to accommodate a request are detailed later. As requests are inserted, the profile is updated to reflect the new resource availability. One node is inserted containing the time at which the request is expected to complete, the number of resource available after its completion, and the ranges of resources available once it completes.

Figure~\ref{fig:tree_example}~(b) illustrates the resulting red-black tree, where shaded circles are black nodes. Each node represents a time and contains the information presented in Figure~\ref{fig:tree_example}~(c), where dashed lines are the linked list connecting sibling nodes. For the profile presented in Figure~\ref{fig:iterating_tree} as an example, to accept a reservation request whose start time is 220, finish time is 700, and requires 2 resources, the algorithm:

\begin{figure}[ht]
\centering 
\includegraphics[width=.95\columnwidth]{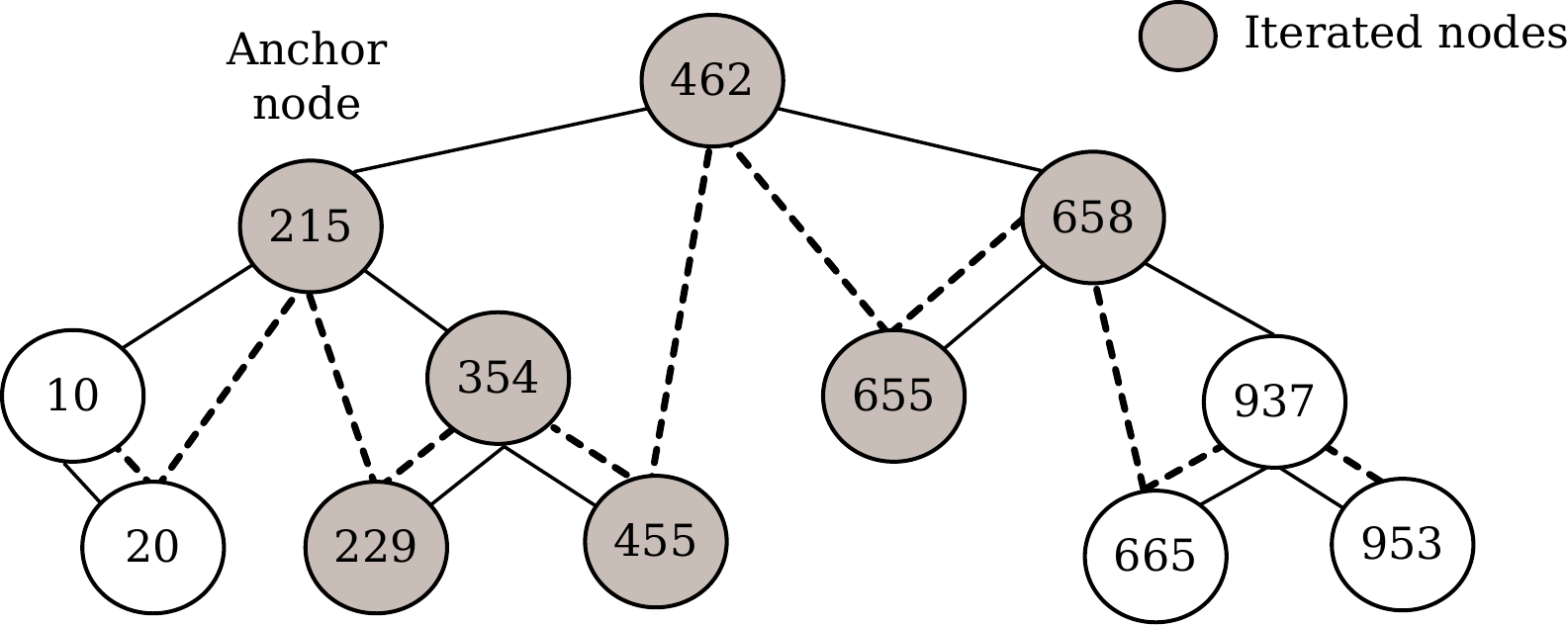} 
\caption{Iterating a tree using the linked list to perform admission control of a reservation request to start at 220 and finish at 700.}
\label{fig:iterating_tree}
\end{figure}

\begin{enumerate}
\item Obtains from the reservation the start time, finish time, and number of resources required.
\item Uses the reservation start time to find the node (\textit{i.e.} the anchor) whose time precedes or is equal to the reservation's start time. If the anchor does not have enough resources, then the request is rejected.
\item Examines the ranges if the anchor has enough resources to serve the reservation. Then, uses the list to iterate the tree and examine all nodes whose times are smaller than the reservation's finish time. For each node, the algorithm computes the intersection of the node's ranges with the intersection of ranges from previously examined nodes. If the the resulting intersection has enough resources to serve the request, then it is accepted.
\item Stops and rejects the request whenever a computed range intersection does not have enough resources.
\end{enumerate}

\begin{figure}[ht]
\centering 
\includegraphics[width=.95\columnwidth]{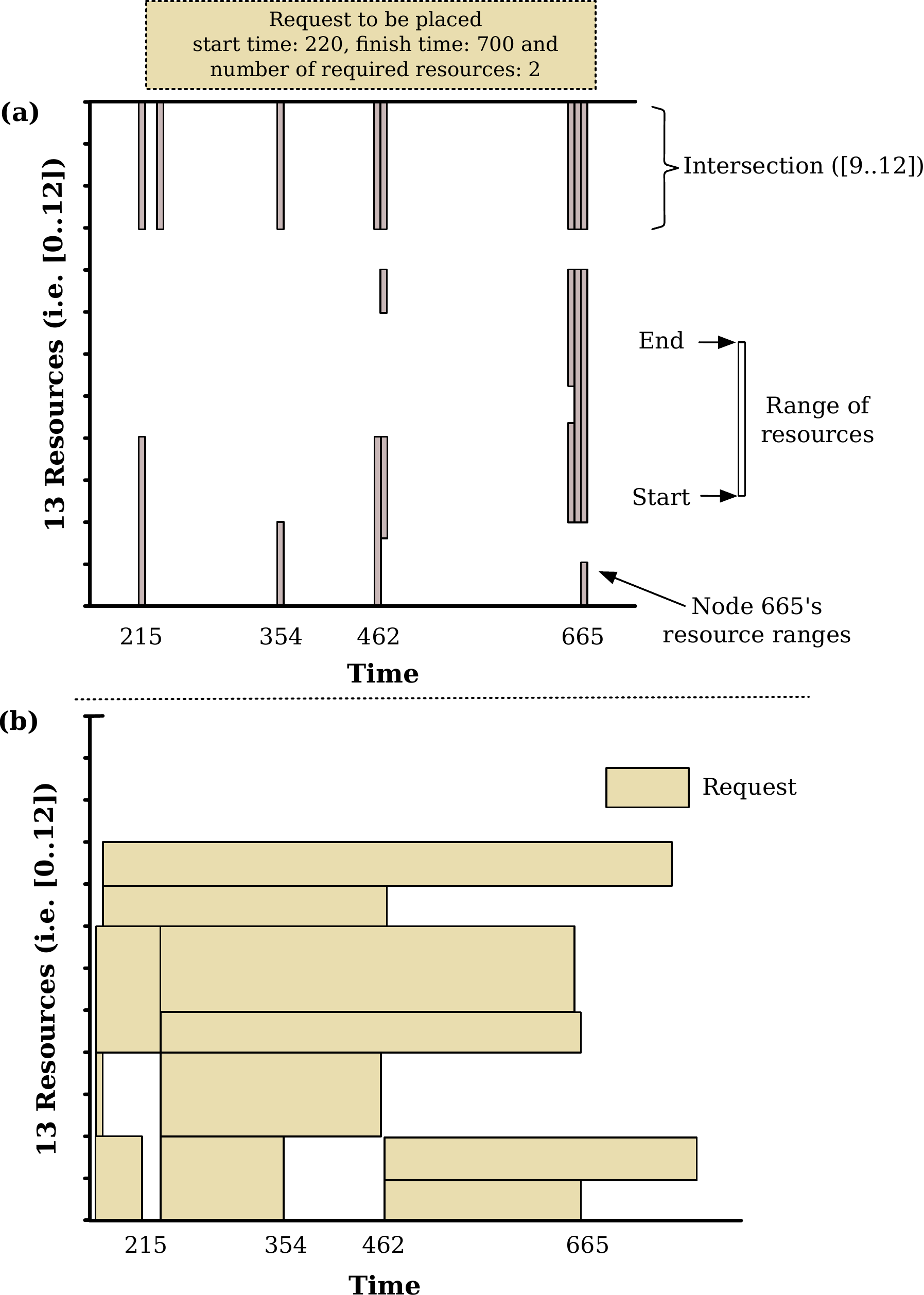} 
\caption{Part of a scheduling queue as (a) ranges of available resources and; (b) reservations. To accommodate a reservation request, the intersection of available ranges must have enough resources.}
\label{fig:ranges_available}
\end{figure}

Figure~\ref{fig:ranges_available} (a) illustrates the relevant part of the profile represented as lists of resource ranges available over time, whereas Figure~\ref{fig:ranges_available} (b) depicts the actual scheduling queue with the corresponding reservations. The new request can be accepted in this case because the intersection of ranges has more resources than what is required.

\subsection{Operations}

The implementation of the availability profile contains several operations to:

\begin{itemize}
\item Check whether a reservation with strict start and finish times can be accommodated.
\item Find a time frame over which a request with flexible start and finish times can execute.
\item Obtain the availability information (\textit{i.e.} free time frames) in the profile.
\item Get the scheduling options for a request or reservation, which are important for schedulers based on strategies such as best-fit and worst-fit.
\item Add time frames to a profile when requests are cancelled or paused, or new resources are added to a pool. 
\item Reconstruct availability profiles from time frames.
\item Allocate time frames to requests.
\end{itemize}
 
Next we describe two operations namely to check resource availability (\textit{i.e.} to decide whether a request can be served) and to update the profile by allocating the resource ranges assigned to the request.

\subsubsection{Check Availability}

As discussed earlier, we consider two types of requests; reservations, that require resources at a well-defined time frame; and best-effort requests that accept resources as they become available.

\IncMargin{-0.6em} 
\RestyleAlgo{ruled}\LinesNumbered
\begin{algorithm}[htp]
\caption{Find a time frame to accommodate a request.}
\label{algo:find_job_slot} 
\DontPrintSemicolon
\SetAlgoLined
\SetAlgoVlined
\footnotesize{
\SetKwInOut{Input}{input} 
\SetKwInOut{Output}{output} 

\Input{the request's duration and number of resources (reqRes)} 
\Output{a profile entry with the request's start time and available ranges} 
\BlankLine

$ctime \leftarrow $ the current time\;
$iter \leftarrow $ profile iterator starting at the node preceeding $ctime$\;
$intersec \leftarrow null$\;
$pstime \leftarrow ctime$ // request's potential start time\;
$pftime \leftarrow -1$ // request's potential finish time\;
$anchor \leftarrow null$\; 

\While{iter has a next element} { 
	$anchor \leftarrow $ the next element of iter\;
	\eIf{$anchor.nRes < reqRes$} {
		\textbf{continue}\;
	}{
		// a potential anchor has been found\;
		$pstime \leftarrow anchor.time$ // potential start is anchor's time\;
		$pftime \leftarrow pstime + duration$ // potential finish time\;
		$intersec \leftarrow anchor.ranges$ // intersections of ranges\;
		$ita \leftarrow $ profile iterator starting after $pstime$\;
		\While{ita has a next element} { 
			$nxnode \leftarrow $ the next element of ita\;
			// does not check nodes beyond potential finish time\;
			\eIf{$nxnode.time \geq pftime$}{
				\textbf{break}\;
			}{
				\If{$nxnode.nRes < reqRes$} {
					// not enough resources available\;
					$intersec \leftarrow null$\;
					\textbf{break}\;
				}
				
				$intersect \leftarrow intersect \cap nxnode.ranges$\;
				\If{$intersec.nRes < reqRes$} {
					// not enough resources available\;
					\textbf{break}\;
				}
			}
 		}
		
		\If{ $intersec.nRes  \geq  reqRes$ } {
			// found time frame with enough resources\;
			\textbf{break}\;
 		}
	}
}

$entry \leftarrow $ new entry with $time=pstime$\;
$entry.ranges \leftarrow intersec$\;
\Return $entry$\; 
}
\end{algorithm}
\IncMargin{0.6em} 

As mentioned earlier, the process of checking whether a reservation request can be accommodated, comprises of first finding the anchor by using the tree, and then iterating the list to check all nodes lying within the anchor and the last node before the requested finish time. The worst-case scenario for checking whether a reservation can be admitted into the system is $O(log\ n + m)$ or $O(m)$, where $log\ n$ is the cost of finding the anchor node in the tree and $m$ is the number of nodes of the sub-list between the anchor and the last node before the request finish time.

To schedule a best-effort request, which can be served at any time, the algorithm can start iterating the tree using the current time as start time. Different from the admission of a reservation, however, to find a time frame for a best-effort request the algorithm starts with a potential anchor; a node with enough resources to serve the request. The intersection of the potential anchor's resource ranges with the following nodes' ranges until before the expected completion of the request needs to have enough resources to accommodate the request. The worst-case scenario for this operation is $O(log\ n + m^2)$ or $O(m^2)$, where $log\ n$ is the cost of finding the first anchor in the tree and $m$ is again the number of nodes of the sub-list between the first potential anchor until the end of the list. The pseudo-code for this procedure is depicted in Algorithm~\ref{algo:find_job_slot}. 

The profile also provides operations for the scheduler to obtain the \textit{free time frames}. A free time frame contains information about the resources available over a given time interval. A time frame has a start time, a finish time, and the ranges of resources available. The availability profile has two operations to obtain the free time frames. The first operation returns free time frames that do not overlap with one another; similar to the approach used by Singh~\textit{et al.}~\cite{SinghHPDC:2007} in their extended conservative backfilling policy. The complexity of this operation is $O(log\ n + m^2)$ or $O(m^2)$ in the worst-case scenario, where $log\ n$ is the cost of finding the the anchor for the query's start time and $m$ is the number of nodes in the list between the anchor node and the last node before the query's end time. In real scenarios, however, this operation is not invoked often, as the operations required to check whether a request can be admitted generally do not need to obtain a list of free time frames.

The second operation to query free time frames returns a list of scheduling options~\cite{JacksonMAUI:2001}, where time frames overlap. The returned time frames are termed as \textit{scheduling options} because they represent positions in the queue where a given request or reservation can be placed. This operation is useful when a scheduler needs to perform more complex selection of resource ranges for a best-effort or reservation request. For example, in some systems the users are allowed to extend previous reservations or resource leases~\cite{InwinShirako:2006}. The scheduler may be required to select a free time frame for a reservation that can accommodate a potential extension or renewal of the resulting resource lease.

\subsubsection{Updating the Profile} 

Once a time frame for a request is found, and it is accepted, the availability profile needs to be updated accordingly, which consists of:

\begin{itemize}
\item Updating the anchor or inserting a new node if the request's start time does not coincide with the anchor's.
\item Updating all entries from the anchor until before the request's completion time, removing the selected resource ranges.
\item Inserting a new node marking the completion time of the request containing the ranges available once the request completes.
\end{itemize}

To minimise the number of nodes in the tree, requests with the same start time or completion time share nodes. The worst-case complexity of the update operation is $O(log\ n + m)$ or $O(m)$ as it consists in inserting one element in the tree (\textit{i.e.} $log\ n$) and updating the $m$ nodes until before the completion of the request, iterating the linked list.

\begin{figure}[ht]
\centering 
\includegraphics[width=0.84\columnwidth]{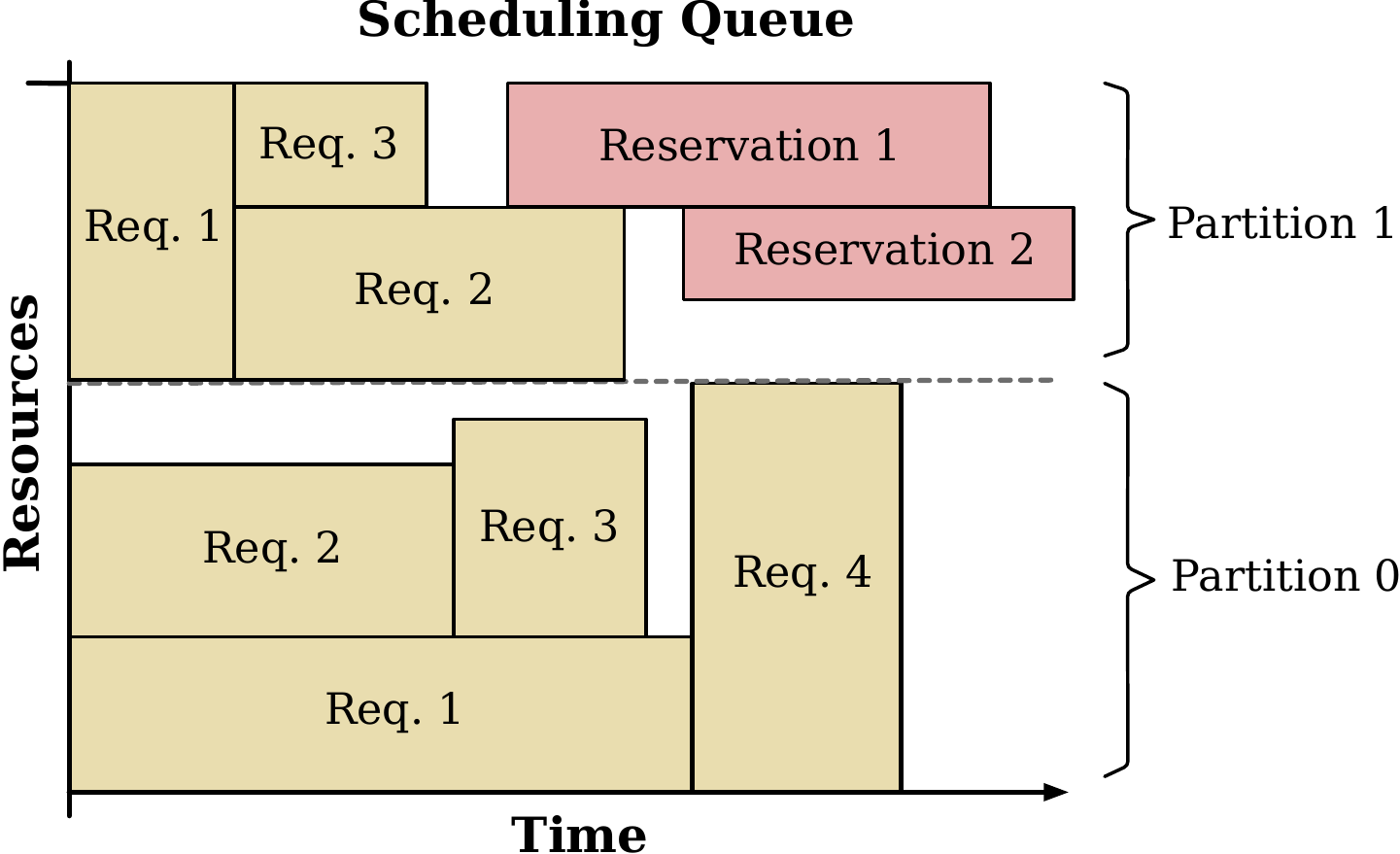} 
\caption{Example of a profile with two resource partitions.}
\label{fig:multiple_partitions}
\end{figure}

\subsection{Multiple Resource Partitions}

An availability profile that controls the allocation of resource ranges to different resource partitions or pools~\cite{LawsonMultiQueue:2002} is also provided. This data structure, termed as \textit{partitioned profile}, is depicted in Figure~\ref{fig:multiple_partitions}, where nodes store the ranges available at more than one resource partition. A user can check the availability of a given partition as well as update that particular partition. As this profile is just an extension of the previously described structure, it is possible to create allocation polices based on the partitioned profile, that allow a partition to borrow resources from another. To enable borrowing, a user uses the operations offered by the normal profile.

\subsection{Implementation Details}

The data structure has been implemented both in Java and Python. An early version has been included in GridSim \cite{BuyyaGridSim:2002}; a grid simulation toolkit that enables modelling and simulation of clusters of computers, grids, storage devices and network topologies. The structure has also been used by resource allocation policies in previous work \cite{AssuncaoHPDC:2009,AssuncaoHiPC:2008} and in schedulers of system prototypes \cite{CostanzoInterCloud:2009}.

\section{Using the Profile}
\label{sec:using_profile}

We show here how to use the profile to build a conservative backfilling scheduler~\cite{MuAlemBackfilling:2001}, but it should be straightforward to implement other policies. The example also demonstrates how to obtain the scheduling options for a request so that a scheduler can select resources using approaches such as best-fit and worst-fit to minimise a queue's fragmentation and improve resource utilisation~\cite{JacksonMAUI:2001,RoblitzElasticAR:2004,RoblitzARPlacement:2006,WieczorekARWorkflow:2006,SinghHPDC:2007}.

\IncMargin{-0.6em}
\RestyleAlgo{ruled}\LinesNumbered
\begin{algorithm}[ht]
\caption{Sample conservative backfilling scheduler.}
\label{algo:conserv_policy} 
\DontPrintSemicolon
\SetAlgoLined
\SetAlgoVlined
\SetKwBlock{Begin}{\vspace{-3mm}}{}
\footnotesize{

\textbf{procedure} reqSubmitted(Req r)\;
\Begin{ 
	$success \leftarrow startReq(r)$\;
	\If{$success = $ \textbf{false}}{
		$success \leftarrow enqueueReq(r)$\;
	}
}

\BlankLine
\textbf{procedure} startReq(Req r)\;
\Begin{ 
	$ctime \leftarrow $ gets current time\;
	$anchor \leftarrow profile.check(j.nRes, ctime, r.duration)$\;
	\eIf{$anchor $ without enough resources} {
		\Return \textbf{false}\;
	}{
		$sls \leftarrow $ select ranges from $anchor$\;
		$profile.allocate(sls, ctime, r.duration)$\;
		$r.ranges \leftarrow sls$\;
		$r.starttime \leftarrow ctime$\;
		\Return \textbf{true}\;
	}
}

\BlankLine
\textbf{procedure} enqueueReq(Req r)\;
\Begin{ 
  // search for an anchor\;
	$anchor \leftarrow profile.check(r.nRes, r.duration)$\; 
	$sls \leftarrow $ select ranges from $anchor$\;
	$profile.allocate(sls, anchor.time, r.duration)$\;
	$r.ranges \leftarrow sls$\;
	$r.starttime \leftarrow anchor.time$\;
}

\BlankLine
\textbf{procedure} resSubmitted(Reserv r)\;
\Begin{ 
	$success \leftarrow admitReserv(r)$\;
	\If{$success = $ \textbf{false}}{
		$options \leftarrow profile.getOptions(r.starttime, r.nRes, r.duration)$\;
		// send scheduling options to user\;
	}
}

\BlankLine
\textbf{procedure} admitReserv(Reserv r)\;
\Begin{ 
	$anchor \leftarrow profile.check(r.nRes, r.starttime, r.duration)$\;
	\eIf{$anchor $ without enough resources} {
		\Return \textbf{false}\;
	}{
		$sls \leftarrow $ select ranges from $anchor$\;
		$profile.allocate(sls, ctime, r.duration)$\;
		$r.ranges \leftarrow sls$\;
		$r.starttime \leftarrow r.starttime$\;
		\Return \textbf{true}\;
	}
}

}
\end{algorithm}
\IncMargin{0.6em} 

Algorithm~\ref{algo:conserv_policy} shows the scheduler main operations where a request is scheduled as it arrives~\cite{MuAlemBackfilling:2001}. Operation \textit{reqSubmitted(Req r)}, called when a request $r$ arrives, tries to start $r$ immediately by calling \textit{startReq(Req r)}. Procedure \textit{startReq(Req r)} executes the same steps required to admit a reservation (\textit{i.e.} \textit{admitReserv(Reserv r)}) as all best-effort requests are initially treated as immediate reservations when they arrive. If a request cannot start immediately, the scheduler finds an anchor with the time at which the request can start; procedure depicted by \textit{enqueueReq(Req r)}. Once the anchor is found, the scheduler selects the resource ranges and updates the profile accordingly.

\section{Experimental Setup and Results}
\label{sec:experiments}

We evaluated the practical and average complexities of two operations, namely (i) assessing the resource availability to find a time frame where a request can be placed (\textit{i.e.} schedule operation), whose theoretic worst case is $O(m^2)$, where $m$ is the number of elements in the sub-list after a first potential start time is found using the RB tree; and (ii) checking whether a given advance reservation can be granted (\textit{i.e.} check operation), whose complexity is $O(m)$ where $m$ is the number of nodes after the start time, as discussed in Section~\ref{sec:availability}.

We used a discrete event simulator developed in house to the model and simulate various scheduling policies using the data structure here. For the experiments reported here, we used conservative backfilling and considered two scenarios, one with a cluster of 1152 CPU cores and another with 446 CPU cores; the latter, in additional to normal requests, permits advance reservations. Although the first scenario does not allow reservations, the scheduler considers a request that arrives initially as a reservation starting immediately, and hence uses the check operation to assess the current resource availability. If the scheduler does not find required resources available, it uses the schedule operation to find a suitable time frame for the request. To drive the workload for the first cluster we obtained one year of request traces (from Jan. to Dec. 2002) from the SDSC Blue Horizon machine\footnote{Parallel Workloads Archive: http://www.cs.huji.ac.il/labs/parallel/workload/}. For the second scenario, we used one year of request logs (from Jan. to Dec. 2013) from four clusters at the Lyon site of Grid'5000 \cite{Grid5000}. We ignored the results of the first and last 4,000 calls to each operation to minimise the impact of warm up and cooldown phases. In both scenarios, more than 60,000 calls to each operation are taken into account.  

\begin{figure}[ht]
\centering 
\includegraphics[width=1.\columnwidth]{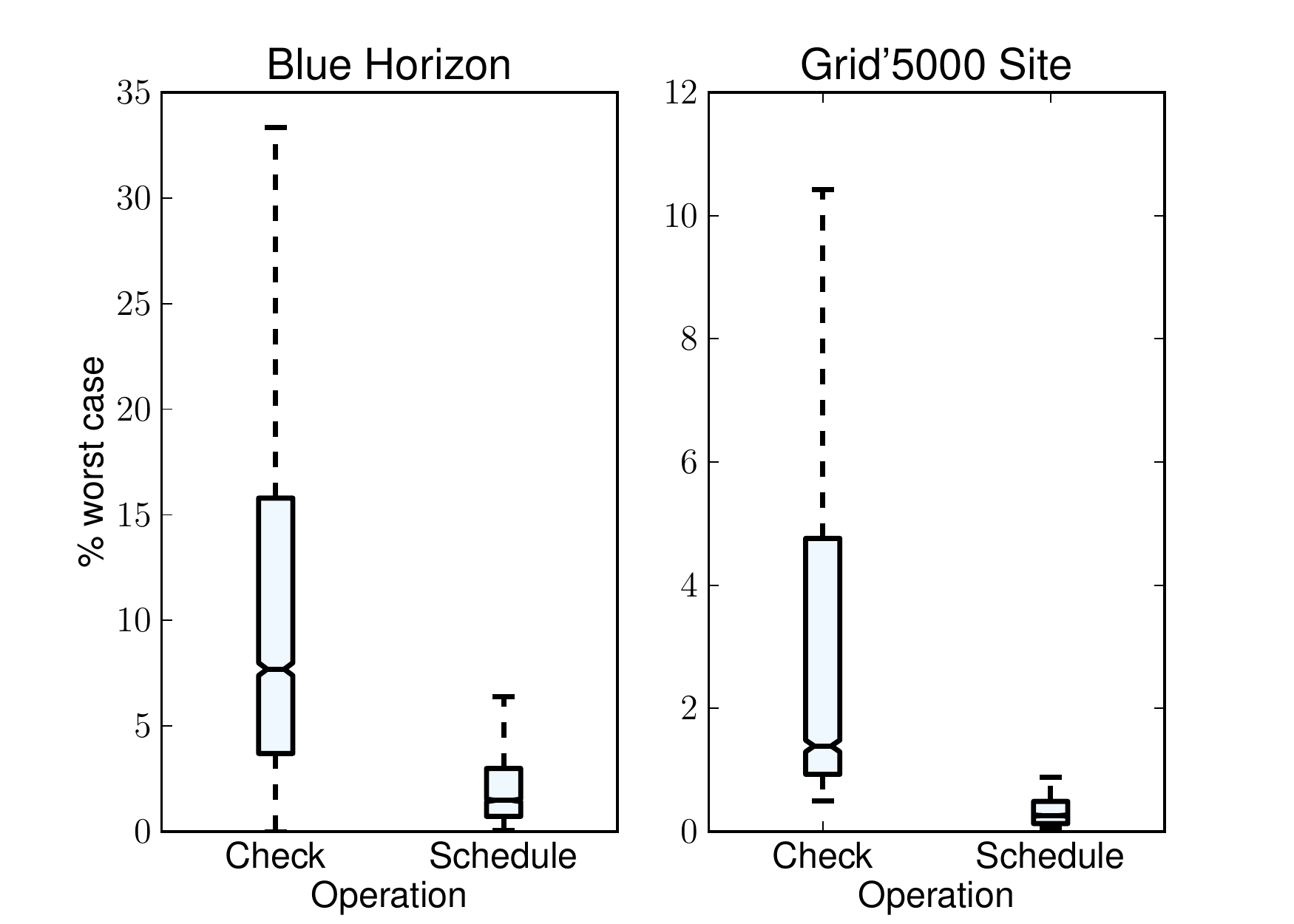} 
\caption{Average complexity of check and schedule operations.}
\label{fig:avg_complexity}
\end{figure}

Figure~\ref{fig:avg_complexity} summarises the average complexity for check and schedule operations by showing the percentage of nodes visited while iterating the list, considering what would be the worst case for each access. Although the average complexity is in general low when compared to the theoretical worst case, it is in general higher for Blue Horizon due to larger fragmentation of its queues. Requests made at the Grid'5000 clusters generally span several hours.

\begin{figure}[ht]
\centering 
\includegraphics[width=1.\columnwidth]{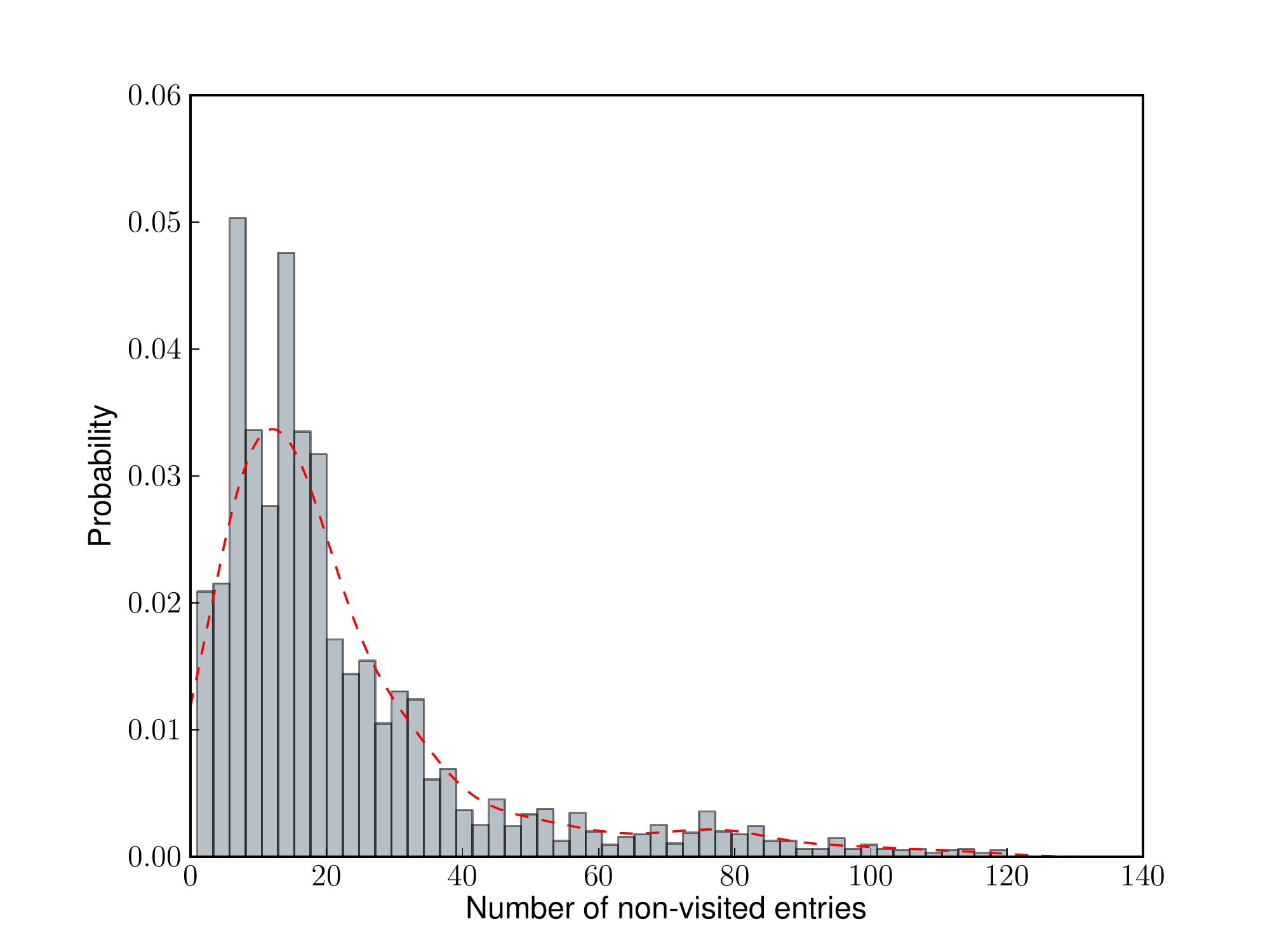} 
\caption{Non-visited entries for the check operation on Blue Horizon.}
\label{fig:sp2_hist}
\end{figure}

Though outliers are not ploted in Figure~\ref{fig:avg_complexity} to improve readability, we noticed that under certain cases the worst case is reached, particularly for the check operation in Blue Horizon. However, after a more detailed inspection, we observed that the worst case is approached when the number of entries to be evaluated is small. For Blue Horizon, Figure~\ref{fig:sp2_hist} shows a histogram of the number of entries that are not visited while the check operation iterates the list.

The histogram shows that the number of non-visited entries is often small, thus demonstrating that even though the average complexity of the check operation is higher for Blue Horizon, the evaluated entry set at each iteration is generally small.  

\section{Conclusion}
\label{sec:conclusions}

This paper presented a data structure to facilitate the scheduling of requests by cloud resource management systems. We provided details about the data structure, which uses a red-black tree to find a potential start time for reservations and a double-linked list to iterate the tree's nodes. We provided an example that demonstrates how the availability profile can be utilised to create scheduling policies and generate alternative offers for advance reservation requests. Experimental results show that the average practical complexity of two operations is often below 10\% of their respective theoretical worst cases.

\section*{Acknowledgments}

\begin{small}
Some experiments were carried out using the Grid'5000 experimental testbed, being developed under the Inria ALADDIN development action with support from CNRS, RENATER and several Universities as well as other funding bodies (see https://www.grid5000.fr).
\end{small}

\bibliographystyle{IEEEtran}
\balance
\bibliography{references}

\end{document}